# New twisted van der Waals fabrication method based on strongly adhesive polymer

Giung Park[1,2,#], Suhan Son[1,2,*,#], Jongchan Kim[1], Yunyeong Chang[3], Kaixuan Zhang[1,2], Miyoung Kim[3], Jieun Lee[1], Je-Geun Park[1,2*]

[1] *Department of Physics and Astronomy, and Institute of Applied Physics, Seoul National University, Seoul 08826, Korea*

[2] *Center for Quantum Materials, Seoul National University, Seoul 08826, Korea*

[3] *Department of Materials Science & Engineering and Research Institute of Advanced Materials, Seoul National University, Seoul 08826, Korea*

[#]*Equal Contribution*

[*] *Corresponding author:* suhanson@umich.edu; jgpark10@snu.ac.kr

**Abstract**

Observations of emergent quantum phases in twisted bilayer graphene prompted a flurry of activities in van-der-Waals (vdW) materials beyond graphene. Most current twisted experiments use a so-called tear-and-stack method using a polymer called PPC. However, despite the clear advantage of the current PPC tear-and-stack method, there are also technical limitations, mainly a limited number of vdW materials that can be studied using this PPC-based method. This technical bottleneck has been preventing further development of the exciting field beyond a few available vdW samples. To overcome this challenge and facilitate future expansion, we developed a new tear-and-stack method using a strongly adhesive polycaprolactone (PCL). With similar angular accuracy, our technology allows fabrication without a capping layer, facilitating surface analysis and ensuring inherently clean interfaces and low operating temperatures. More importantly, it can be applied to many other vdW materials that have remained inaccessible with the PPC-based method. We present our results on twist homostructures made with a wide choice of vdW materials – from two well-studied vdW materials (graphene and $MoS_2$) to the first-ever demonstrations of other vdW materials ($NbSe_2$, $NiPS_3$, and $Fe_3GeTe_2$). Therefore, our new technique will help expand moiré physics beyond few selected vdW materials and open up more exciting developments.

1. **Introduction**

Two-dimensional materials based on van der Waals (vdW) atomic crystals and moiré superlattices have recently attracted enormous attention. For example, several exciting discoveries have been made in vdW heterostructures or homostructures: those new observations include superconductivity, correlated insulating states, and ferromagnetism [1–10].

However, these intriguing quantum phenomena depend strongly on a few technical details of fabrication methods. Two primary conditions are essential to observe these phenomena: a precise control of twisted angle and a clean interface between the top and bottom layers [11]. For example, it is well known that the twisted bilayer graphene physics is very sensitive to the twisted angle, so even a slight rotation of 0.2° from the target angle can be detrimental to achieving superconductivity [4]. A standard method used to fabricate twisted homostructures is the polypropylene carbonate (PPC) tear-and-stack method or cut-and-stack method [12,13]. This method allows precise angular control while maintaining a clean interface [2–4,9,14] by dividing a vdW flake into two pieces (tear) and assembling with the designed twisted angle (stack). This way, a twisted angle can be achieved with greater precision because this method uses the same single flake with the same orientation angle. Furthermore, this method can also naturally ensure a clean interface [11,15].

Although the current PPC tear-and-stack method has proven popular for moiré superlattices with graphene and $MoS_2$, there are two main obstacles to the future development of moiré physics when one tries to work with many other vdW materials. First, the current PPC tear-and-stack method relies solely on the adhesive force between h-BN and target materials. The problem is the weak adhesion of h-BN to some materials and the prerequisite of an insulating top h-BN layer [12,15]. Therefore, materials having weaker adhesive with h-BN are challenging to fabricate using the PPC-based method.

The second issue is the role of the h-BN layer. When h-BN and target sample are stacked at a low angle or subjected to thermal annealing, unexpected moiré phases can be generated due to effects such as self-rotation [16–19]. The resulting moiré cell interaction can significantly affect the material's properties [20–22]. To mitigate this effect, special attention must be paid to the angle between the h-BN capping layer and the target layers during fabrication to ensure that the undesired effect is minimized or avoided altogether. Another caution is needed relating to controlling the top h-BN layer when measuring the surface of twisted samples directly. For instance, it is very challenging to use scanning probe microscopic techniques (SPM) that generally require direct contact between the tip and the surface [23,24].

Here, we report a novel tear-and-stack approach for moiré superlattices based on the strongly adhesive polymer polycaprolactone (PCL) [25]. We demonstrate that it is possible to fabricate twisted homostructures without h-BN top layers. Our new method enables many new combinations of twisted homostructures using various sets of vdW materials: for example, graphene, superconducting $NbSe_2$, magnetic $NiPS_3$, and $Fe_3GeTe_2$. Another advantage of our method is that it can be done without vdW capping layers and at relatively low working temperatures. We successfully fabricate several new devices using our method and further demonstrate that the quality of our twisted homostructures is comparable to those made using the PPC-based method. To further expand the applicability of our new method, various protecting layers (such as $ZnPS_3$, h-BN, $CrPS_4$) were successfully tested to build homostructures of graphene and $MoS_2$, as well as $NbSe_2$, $NiPS_3$, and $Fe_3GeTe_2$, which have been challenging, if not impossible. To assess the quality of the samples produced using the new PCL-based method, the interfaces were examined using scanning transmission electron microscopy (STEM) and atomic force microscopy (AFM). Furthermore, Raman spectroscopy and vertical Josephson junction (vJJ) measurements were carried out on twisted bilayer $MoS_2$ and few-layer superconducting $NbSe_2$, respectively.

## 2. Method

### 2.1 PCL stamp and materials preparation

In all experiments conducted in this study, we used a PCL solution with a concentration of 15 % in tetrahydrofuran (THF) from Sigma Aldrich (average Mn: 80000). We first attached a PDMS stamp to a prepared slide glass and then used transparent tape (Scotch® Crystal Tape) to fix it. Subsequently, the PCL polymer was spin-coated onto the stamp at 2000 rpm for 20 seconds. Afterwards, the PCL stamp was annealed for 10 minutes at around 50-70 °C to achieve a flat surface.

We synthesized all our vdW samples: $NiPS_3$, $NbSe_2$, $Fe_3GeTe_2$, $CrPS_4$, and $ZnPS_3$ crystals, using the chemical vapor transport (CVT) method following the recipe provided in references [25–28]. We also used commercial samples: multilayer graphene on $SiO_2$ (285 nm / Si substrate (Graphene supermarket), graphene, $MoS_2$, and h-BN crystals (HQ graphene). Using the mechanical exfoliation method, all crystals were exfoliated onto a $SiO_2$ (285 nm)/ Si substrate. All fabrication processes, including exfoliation, were carried out inside a glove box ($O_2 \leq 0.5$ ppm, $H_2 \sim 0$ ppm) to obtain a clean interface.

### 2.2 AFM and STEM measurement

Using the NX10 AFM (Park Systems), a contact mode tip was utilized to determine the height of our samples, and we confirmed the cleanliness of the surface of the vdW materials. $NbSe_2$ and $MoS_2$ were also prepared with a focused ion beam (FIB) milling for STEM imaging to check the interface. Annular dark-field STEM images were obtained with spherical aberration-corrected STEM (Themis Z, Thermofisher), and STEM images were taken from Themis operated at 80kV with beam convergence semi-angle 25.1 mrad and ADF collection angle 65-200 mrad.

### 2.3 Raman measurement

Raman spectra were measured under ambient conditions at room temperature using a 532 nm continuous laser (Cobolt laser) with a laser power of less than 300 µW to prevent damage to the samples. A 100x objective lens with a numerical aperture of 0.8 was used for laser excitation and signal collection. Raman signals were resolved by a spectrometer with a 1200 grooves/mm grating (SpectraPro HRS-300, Princeton Instruments) and detected by a charge-coupled device. Three 532 nm notch filters (OptiGrate) were employed to eliminate the laser line. Furthermore, all Raman spectra measured on samples were subtracted by spectra taken at adjacent bare substrate sites to remove any residual laser line not eliminated by the notch filters.

### 2.4 Transport measurement

We put our samples onto a $SiO_2$ (285 nm)/Si substrate for transport measurements. Electrodes were patterned on the vJJ structure using e-beam lithography, followed by evaporation of Ti/Au (5/80 nm). The electrodes were fabricated within an hour of taking the samples out of the glove box to prevent oxidation, and all processes were carried out at room temperature.

## 3. Results and discussion

### 3.1 The principle of method

In the current moiré physics, high angular control is achieved using the standard method of separating a single flake into two flakes for the top and bottom parts. Among the existing methods, the PPC tear-and-stack method is the most common approach to treating single flakes [12]. This method is carried out in the following order. The surface of the peeled target sample is held on the crystalline face of h-BN using the attached stamp before varying the temperature. If the adhesion between the target sample and h-BN is stronger than with the substrate, the part attached to h-BN can be lifted off while the other part remains on the substrate. Then, by adjusting the stamp/h-BN/target sample angle using the transfer stage, a twisted sample capped with h-BN can be fabricated onto the sample still remaining on the substrate. This method allows a good separation of the target sample regions via h-BN while maintaining a clean interface and enabling precise angular adjustment.

Unfortunately, for most materials, the adhesion of the target materials is stronger with the substrate than with the h-BN. Therefore, this method can only be applied to a few selected materials, such as graphene, certain TMDCs, and $CrI_3$. This technical limitation of the PPC tear-and-stack method motivated us to develop a new approach extending to a much wider range of vdW materials, particularly many magnetic vdW systems. We use the non-sticking characteristic of certain van der Waals material combinations to develop a novel tear-and-stack method. Thanks to the strong adhesive force of PCL, the number of possible combinations for twisted homostructure increases dramatically, which could eventually lead to an entirely novel class of new twisted homostructure.

Using a micromanipulator installed inside a glove box, we aligned the PCL stamp on top of the protecting layer (the green flake in Figure 1 (a)). In this case, the protecting layer can be any material not sticking to the target layer. We then increased the adhesive force of the PCL stamp by raising the temperature to 55 °C while maintaining contact with the protecting layer. After inserting the protecting layer, we slowly raised the temperature to 63 °C before lowering it to enhance the adhesion strength between the PCL stamp and the protecting layer. Therefore, our method works at relatively low working temperatures compared to the PPC-based method, which will be advantageous for samples fragile at higher temperatures. Note that the strong adhesive force of PCL allows for a wide variety of protecting layers to choose from [25].

We prepared the target sample after picking up the protecting layer. Then, we carefully aligned the protecting layer/PCL stamp on the area to be protected, as shown in Figure 1 (b), before assembling them at 63 °C. In our case, the boundary of the protecting layer serves as a nice cutting line. After waiting for 1-2 minutes to sufficiently strengthen the adhesion between the PCL stamp and the target layer, we carried out the tear process at 30 °C, as shown in Figure 1 (c-d). In our method, the protecting layer must have a vdW force weaker than that between the target layer and the substrate, allowing only the portion in direct contact with the PCL stamp to be torn and lifted off. Given the weak interlayer attraction between most vdW materials, selecting a protecting layer poses no problem for our method. To achieve a more precise evaluation of the suitability of materials, it is essential to measure the vdW forces between each vdW material quantitatively. Therefore, direct measurements of interfacial adhesion can provide a better guide about material choice for the target material [29–31].

Furthermore, as depicted in Figure 1(d), the protecting layer can effectively replace h-BN in determining the cut line for the PPC tear-and-stack method. This feature allows one to make conveniently twisted samples with desired shapes. Here, the different adhesive forces between two regions of the vdW flake led to the successful tearing process. This tearing process can also be assisted by other novel methods, especially in the case of the vdW flake with strong covalent bonds [32]. Figure 1 (e) shows that the PCL stamp or substrate can be rotated and placed at the desired angle. Finally, removing and cleaning the PCL stamp can result in a twisted homostructure, as shown in Figure 1(f).

There are two specific points to be noted about our method. (1) First, the interface between twisted homostructure can, in principle, be free of polymer residue during the entire process. (2) Second, the finally fabricated twisted homostructure (shown in Figure 1(f)) does not necessarily need a capping layer on the twisted region. Therefore, the first aspect of our method naturally guarantees a clean interface, which is possible because of the protecting layer (the green flake). Additionally, the second aspect gives new opportunities for surface-sensitive measurements with vdW twist samples. Note that an h-BN capping layer must always be the topmost layer in the conventional tear-and-stack method, making direct contact challenging between the probing tips and the surface of the samples. Therefore, the fact that our method can be used without a capping layer is a clear advantage for surface-sensitive studies such as scanning tunnelling microscope (STM), magnetic force microscope (MFM), and AFM. It would also avoid unnecessary potential surface modifications resulting from the lattice mismatch between the capping and target layers. Table S1 in Supplementary Information compare our method with the other existing methods more directly.

To demonstrate this point, Figure 2(a) shows that the surface of the twisted homostructures of our device can be directly observed with an AFM tip. This figure clearly illustrates that the protecting layer (depicted in yellow) does not cause problems with the AFM measurements as it can be easily positioned adjacent to the twisted region. Moreover, by modifying the cut line through the protecting layer, the pristine region can be transformed into a non-capped region, enabling simultaneous surface investigations alongside the twisted part. To further illustrate its broad applicability with other vdW materials, we tested our method using various materials, including the two well-known materials of graphene and $MoS_2$. Another significant advantage of our method is that we can work with larger-scale twisted samples than have been possible. As a demonstration, we fabricated a large-area twisted homostructure using bilayer $MoS_2$ and multilayer Graphene (See Supplementary Information Note 4). This particular aspect of our method will open up another uncharted area of twisted physics.

### 3.2 Clean van der Waals interface

The clean interface of the vdW heterostructure is essential to minimize the external and extrinsic effects. It can easily contaminate the interface if the vdW interface is exposed to the polymer residue during the heterostructure assembly or the lithography process. For instance, the polymer residue sensitively affects the quality of electrical contact or makes bubbles and wrinkles at the vdW interfaces [33]. On the other hand, the vdW heterostructure device with a cleaner interface is expected to show much-improved mobility and device quality in graphene devices [34].

In our device fabrication process based on the PCL tear-and-stack method, the interface has no direct contact with the polymer. Therefore, it can naturally avoid possible contamination from any polymer residue. Furthermore, it has another advantage: PCL's strong adhesive force enables novel combinations of twisted homostructures using materials that have been otherwise impossible to use. To demonstrate these points, we prepared and investigated the twisted 5-layer $NiPS_3$ homostructure (Figure 2(b,c) and few-layer $NbSe_2$ homostructure (Figure 2(d)), respectively. The original flakes were torn away and assembled following the process described in Figure 1. Note that to the best of our knowledge, these material combinations have been challenging to adapt to the PPC tear-and-stack method. The AFM image of our final product in Figure 2(c) shows no apparent signs of contamination, bubbles, or wrinkles. As can be seen in the inset AFM data (Figure 2(c)), the thicknesses of the original and twisted regions were measured to be 3 and 6 nm, respectively. Considering that the thickness of the $NiPS_3$ monolayer is about 0.6 nm, this shows that the size of the residue buried in the top layer can be neglected down to the picometer scale. Here, note that those fabricated homostructures are also capping-layer-free. The lack of capping layers in our method is helpful for surface-sensitive tools, which makes this technique a valuable toolkit for the observations of surface-sensitive physics [35–39]. Therefore, our capping-layer-free technique is expected to facilitate surface-sensitive moiré physics, such as twisted-angle-induced ferroelectricity [40].

Furthermore, STEM measurement was also carried out for more direct observation of the interface's cleanliness. Measurements were performed on 30°-twisted NbSe$_2$ and 0.2°-twisted MoS$_2$ (see Supplementary Information for MoS$_2$ data). Figure 2(e) directly demonstrates the absence of any contaminated interfaces. Therefore, we can summarise three major characteristics of our PCL tear and stack method: i) the strong polymer PCL allows a variety of twisted homostructure combinations; ii) it guarantees a clean interface; iii) it does not need a capping layer on the topmost layer.

### 3.3 Moiré phonon in twisted MoS$_2$

In the case of homostructures, interlayer coupling can be easily controlled through twisting. In particular, the twisted angle plays a crucial role in tuning the electronic band structure [41–44], and lattice reconstruction or zone folding effect depends on the twist angle [45,46]. Only a few techniques are available for the existing methods to investigate the effect of twist angle for fabricated twisted homostructures [45,47–49]. Among these techniques, the moiré phonon measurement is quite helpful as it can determine the angle by analyzing Raman shift values within homostructures formed at low and high angles. Because these signals arise due to interlayer coupling perturbations, this approach can also serve as a means to demonstrate the quality of the samples [45,50,51].

To demonstrate the new fabrication method's precision and illustrate the detectability of signals sensitive to the interface, we fabricated twisted-1L MoS$_2$ samples using our new PCL-based method. Using our device, we measured moiré phonons using room-temperature Raman spectroscopy. Due to the D$_{3h}$ (3-fold symmetry) symmetry of MoS$_2$, the Raman peaks shift due to twisting angles is symmetric around 60°, covering the range from 0° to 120° [52]. Additionally, as the moiré reciprocal lattice vector aligns with the 1$^{st}$ Brillouin zone through the monolayer reciprocal lattice vector, there is a 30° symmetry between 0° and 60° [45].

To further examine the zone folding effect in twisted MoS$_2$, several twisted samples were fabricated with eight different twist angles: 0.2°, 10°, 15°, 18°, 25°, 30°, 40°, and 50°. We note that using h-BN as a protecting layer for the PPC tear-and-stack method is challenging due to the strong attraction between h-BN and MoS$_2$ [12]. Thus, our device employed a new protecting layer of ZnPS$_3$, which exhibited no adhesion to the sample: ZnPS$_3$ belongs to the class of well-studied TMPS3 with a large band gap of 3.32 eV [25]. Figure 1(c-d) shows that our method does not need adhesive force to secure the substrate's bottom layer. We comment that ZnPS$_3$ formed in shiny transparent plates is stable in the air, and there is no visible sign of degradation even after a few weeks of exposure to the air. Thus, no extra capping layer was added to our device to manufacture the MoS$_2$ homostructures.

Note that moiré phonon peaks are visibly observed in our sample between 60 and 450 cm$^{-1}$, which varies depending on the twisted angle (Figure 3(a, b). These angle-dependent Raman signals are due to the zone folding effect, which can be compared with the reported phonon dispersion of TMDC's monolayer. Upon close inspection, these peak position changes seen in our device show excellent agreement with the reported phonon dispersion mode of MoS$_2$ monolayers (Figure 3(c)) [45]. Given the excellent alignment between the two data sets, our PCL tear and stack method can easily adjust the angle to ensure a clean interface. In addition, we confirmed the stiffness of the peaks depending on the angle below 100 cm$^{-1}$, as in Figure 3(d). While interlayer shear (S) mode and interlayer breathing (LB) mode were observed in the 0.2° twisted MoS$_2$ and 2L MoS$_2$, only the LB mode was observed in other twisted samples. This can be explained by competition between strain and interlayer coupling [46]. Using this, we can categorize angles into three regimes: a relaxed regime ($\theta < 2°$), a transition regime ($2° < \theta < 6°$), and a rigid regime ($\theta > 6°$) [46,53]. We can also see that both an untwisted bilayer MoS$_2$ and a pristine bilayer MoS$_2$ follow the same relaxed regime, showing both S and LB modes. Other samples follow the rigid regime, showing only the LB mode because all twisted angles are over 6°. These results represent a clear example of the moiré effect beautifully demonstrated in our device.

### 3.4 Josephson junction in twisted NbSe$_2$

To further demonstrate our method's clean interface and new application, we fabricated twisted NbSe$_2$ homostructures and measured the critical current density: the weak link between two NbSe$_2$ superconductors form the Josephson junction for various reasons [54–57]. For our purpose, 2H-NbSe$_2$ is particularly useful for its superconducting properties and various physical phenomena. It is also important for our attempt that this combination of excellent physical properties has made 2H-NbSe$_2$ an interesting choice for studying and exploring the unique characteristics and behavior of vdW vJJ. [58–61]. Therefore, 2H-NbSe$_2$ can be an excellent benchmark test. We were also motivated by the fact that 2H-NbSe$_2$ is known to be difficult to fabricate as twisted homostructures with the existing PPC tear-and-stack method [25]. Another interesting point is that both the critical temperature and the superconducting gap size depend on the layer number [62], with an anticipation that these modifications result in more significant fluctuations in physical properties as we approach the two-dimensional threshold [59]. To address this issue, therefore, it is essential to ensure uniform thickness for both the top and bottom layers, which seems to be a formidable challenge for the PPC-based method [55].

However, our new method allows us to fabricate the first-ever equal-thickness twisted NbSe$_2$ device while maintaining the precise twisted angle and clean interface. For NbSe$_2$, the superconducting gap or critical temperature changes significantly as the bulk limit approaches the two-dimensional limit, so the target layers were set to samples with more than 10 layers [62]. Using this new device, we examined the transport characteristics of the vertically stacked NbSe$_2$-NbSe$_2$ Josephson junction. Figure 4(a) shows that the resistance drops to zero within the junction region at T$_C$ ~7.0 K, the same as the bulk NbSe$_2$ (~7.2 K). Our device has the RRR (residual resistance ratio) value of approximately ~11, defined as RRR=R(T=300 K)/R(T=10 K). Note that this RRR value is similar to the literature [56]. We also observed no broadening near the transition temperature. When considered together with the benefits of the tear and stack method, this suggests that our device has a clean interface with a vdW gap free of impurities. The temperature dependence of the critical current I$_C$ was calculated from the Ambegaokar-Baratoff (AB) formula using the Bardeen-Cooper-Schrieffer (BCS) theory [57]. Our analysis of the critical current confirms the $\Delta(0)$ of our junction is about 1.11 meV, as shown in Figure 4(b).

With these insights, we investigated the angular dependence of the critical current density at different angles, as illustrated in Figure 4(c). The effect of twisted angles on the critical current density in superconducting twisted homostructures has been well-documented in the literature [56]. And a similar angular dependence has also been reported in devices with clean and flat interfaces [63]. In our experimental setup, the current channel within the k-valley of NbSe$_2$ is expected to produce a notable effect on the junction current due to its k-dependent s-wave superconductivity [56]. Interestingly, although two NbSe$_2$ samples with twist angles of 0° and 60° exhibit nearly equivalent critical current densities (~3000 A/cm$^2$), a symmetrical point exists at around 30°. It also occurs when the distance between the k-valleys of the upper and lower layers is maximized above and below the interface. This specific minimum value is additional evidence corroborating the presence of a k-valley superconducting gap in NbSe$_2$. Moreover, these features appear irrespective of the NbSe$_2$ layer count in bulk scenarios, as referenced [64].

We demonstrated all our results are highly reproducible using our method. It is also to be noted that the quality of our device is comparable to the previous results [56]. Moreover, our method has other distinct advantages compared to the previously fabricated homostructures like the PDMS method, which are not based on the tear-and-stack method [56,63,65]. This method inevitably leads to different thicknesses for the top and bottom layers. In contrast, our PCL tear-and-stack method naturally ensures the same thickness for the top and bottom layers with good control of orientation and twisted angle. We believe that this flexibility of our method will allow exciting, otherwise tricky experiments, including fabricating novel twisted homostructure combinations.

## 4. Conclusion

In summary, we report a new fabrication method for twisted vdW samples by combining two techniques: PCL stamp and tear-and-stack method. This new PCL-based method offers several advantages: most notably, a broad applicability across many different vdW materials, a flexible choice of protecting layers, and low working temperatures. Our method enables direct contact measurements while maintaining the benefits of the conventional tear-and-stack method and sustaining fine angle control. Additionally, this new method allows for the production of twisted samples in various sizes up to 1 centimeter, irrespective of the capping layer size. Our AFM and STEM measurements confirm the clean surface and interfaces of the fabricated samples. Our method's capping-layer-free nature and the material selection's expandability have great potential to unveil novel observations. This new method is expected to open up an exciting new window of opportunities and further accelerate the already exciting research field.

**ACKNOWLEDGEMENT**

We acknowledge Sungmin Lee, Young Jae Shin, and Philip Kim for their contributions to this project's initial state. We also thank Youngwoo Son and Gil-Ho Lee for their helpful discussions and comments. This work was supported by the Leading Researcher Program of the National Research Foundation of Korea (Grant No. 2020R1A3B2079375). The work by J.K. and J.L. was supported by the National Research Foundation (NRF) of Korea (Grant No. 2020R1A2C201133414).

**Data availability statement**

The data that support the findings of this study are available upon request from the authors.

**Figures**

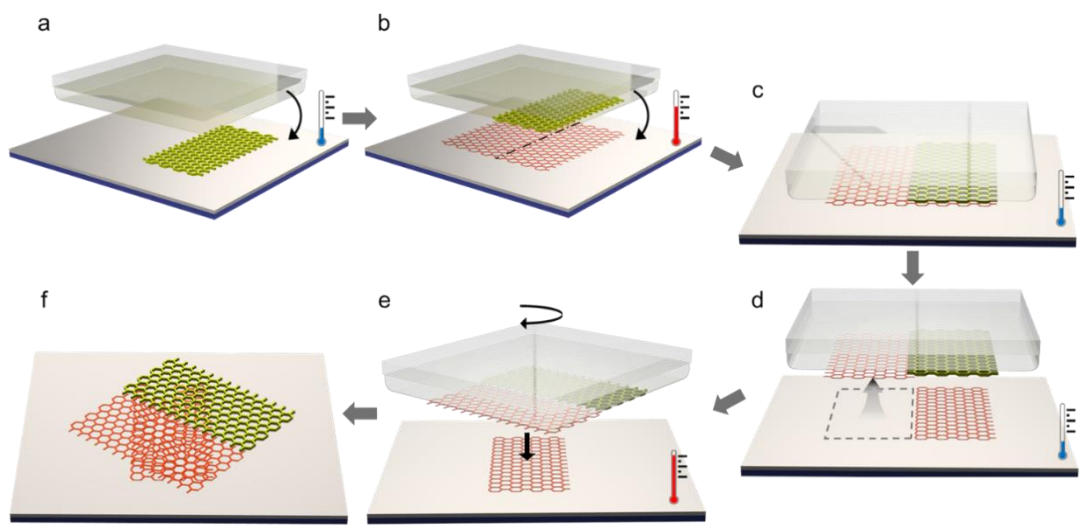

**Figure 1.** Schematics of the PCL tear-and-stack method. (a) Protecting layer picked up by PCL stamp. (b) Finding and aligning the target sample with the PCL stamp. (c) Decreasing temperature increases the adhesive force between the PCL stamp and the target sample. (d) A step of tearing the target sample in the direct contact region by PCL adhesive force while protecting another part. (e-f) Rotating and aligning the upper layer to the bottom layer.

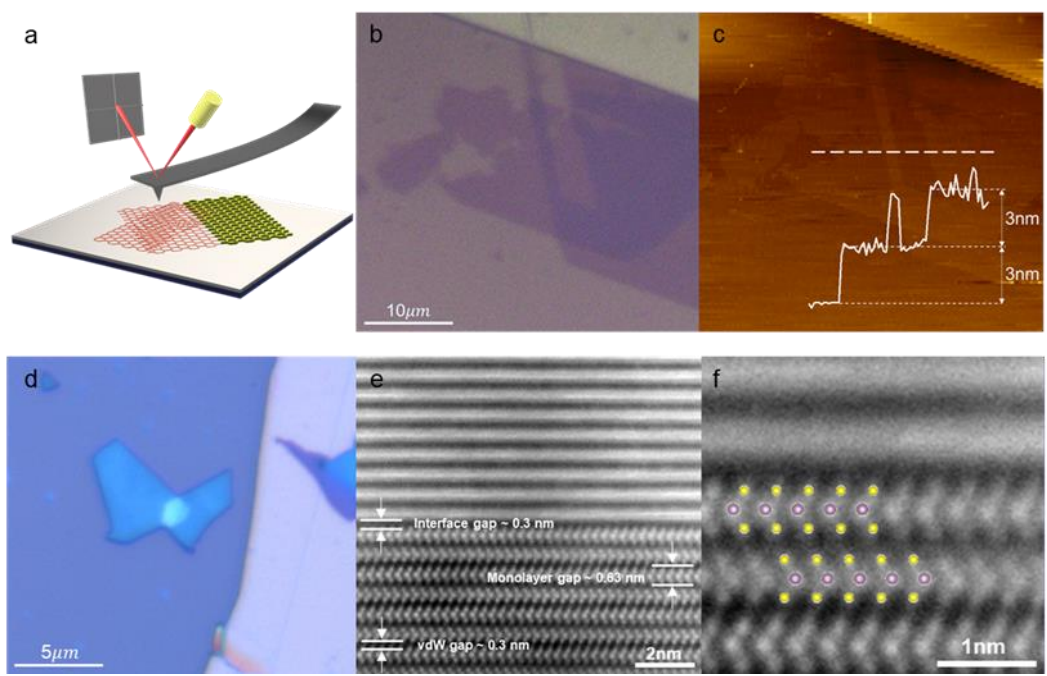

**Figure 2.** A clean interface is demonstrated by the photos taken by AFM and STEM. (a) A schematic diagram for how to make direct tip contact with twisted samples. (b) and (d) The optical micrographs of twisted samples using NiPS$_3$ and NbSe$_2$. (c) The AFM data of twisted NiPS$_3$ with a contact mode. The inset shows a cut line with the raw data. (e-f) A cross-section of STEM data depicting the interface of twisted NbSe$_2$.

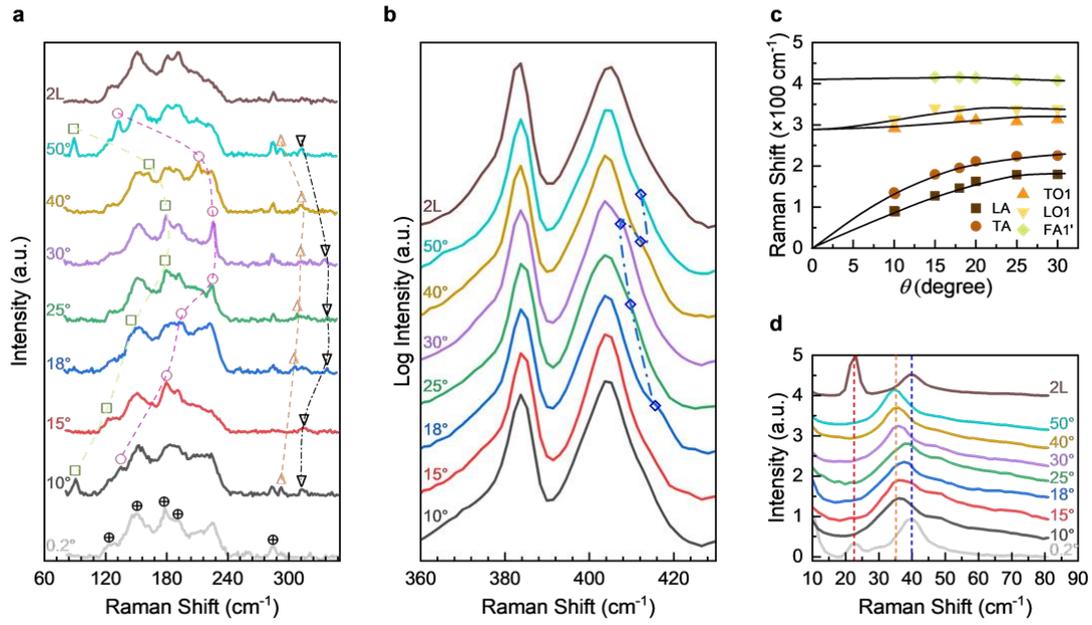

**Figure 3.** Twisted MoS$_2$ Raman data and angular dependence. The Raman spectra of twisted bilayer MoS$_2$ with different angles in the region of (a) 60-360 cm$^{-1}$ and (b) 360-430 cm$^{-1}$. Different shapes and color symbols represent the Raman modes of separate phonon branches. (c) The phonon-dispersion relation with different angles. (d) Interlayer phonon modes are taken with different angles. The guidelines refer to S mode(red) and LB mode(orange, blue) with the excitation photon energy of 2.33 eV (531.97 nm).

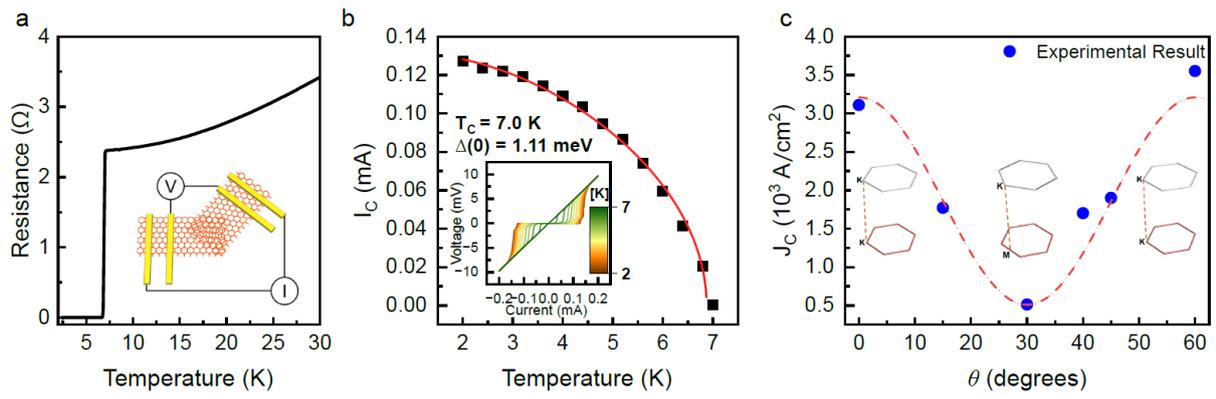

**Figure 4.** Twisted NbSe$_2$ transport data and angle dependence. (a) Temperature-dependence of the resistance for NbSe$_2$ Junction. (sample S2, θ = 15°) (b) Critical current–temperature characteristics of a NbSe$_2$ vertical Josephson junction between T =2 and 7 K. The inset curve represents the I-V curve as a function of temperature. (c) The critical current density variation with twist angles of 0~60° was taken at T=2 K. The dashed line guides the eye.

# Supplementary Information: New twisted van der Waals fabrication method based on strongly adhesive polymer

Giung Park[1,2,#], Suhan Son[1,2,*,#], Jongchan Kim[1], Yunyeong Chang[3], Kaixuan Zhang[1,2], Miyoung Kim[3], Jieun Lee[1], Je-Geun Park[1,2*]

[1] *Department of Physics and Astronomy, and Institute of Applied Physics, Seoul National University, Seoul 08826, Korea*

[2] *Center for Quantum Materials, Seoul National University, Seoul 08826, Korea*

[3] *Department of Materials Science & Engineering and Research Institute of Advanced Materials, Seoul National University, Seoul 08826, Korea*

[#]*Equal Contribution*

[*]*Corresponding author:* suhanson@umich.edu; jgpark10@snu.ac.kr

Contents:

Supplementary Note 1: PCL tear and stack process with real materials

Supplementary Note 2: Additional Cross-sectional STEM data

Supplementary Note 3: Sample images used in measurements.

Supplementary Note 4: Large-scale twisted homostructure

Table S1: Comparison with other techniques.

Figure S1: Schematics of PCL tear and stack method with real materials.

Figure S2: Other van der Waals twisted samples

Figure S3: Twisted $MoS_2$ STEM data

Figure S4: Various twisted $MoS_2$ samples

Figure S5: Various twisted $NbSe_2$ samples

Figure S6: Large scale twisted $MoS_2$ sample

Figure S7: Large-scale twisted Graphene sample

**Supplementary Note 1: PCL tear and stack process with real materials**

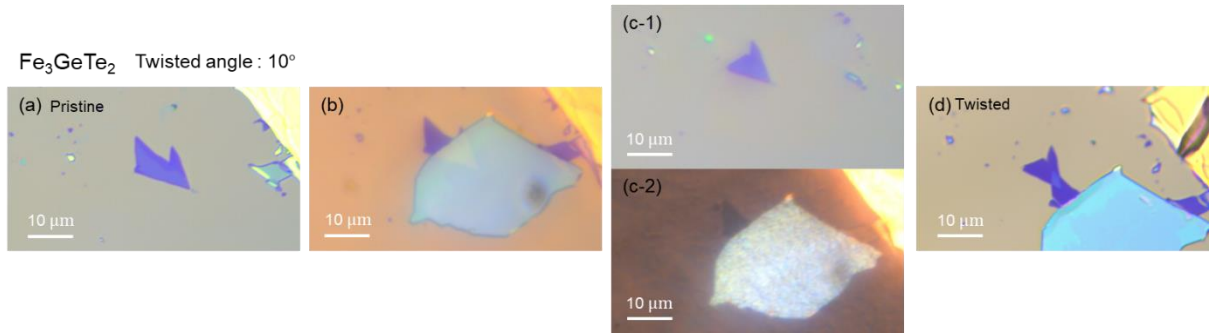

**Figure S1.** (a-d) Optical micrographs of successive steps: $Fe_3GeTe_2$ with a protecting h-BN. (a) Finding target samples. (b) Aligning and attaching the protecting layer to the target layer. (c) Result of tear process. (c-1) Bottom layer on $SiO_2$ substrate, and (c-2) Top layer with PCL stamp. (d) Twist angle and stack process. The twisting angle is 10 °. The scale bar is 10 μm.

| | Our work (PCL tear and stack method) | Polydimethyl siloxane (PDMS) transfer method | Polypropylene carbonate ((P)PC) tear and stack method | Wet transfer method | CVD growth method |
|---|---|---|---|---|---|
| Capping layer free | O | O | X | O | O |
| Precise controllability of rotational angle | ***** | *** | ***** | *** | * |
| Cleanness of interface | ***** | *** | ***** | * | **** |
| Fabrication temperature | Low (< 75 ºC) | Low (< 100 ºC) | High (> 130 ºC) | Very low (Room temperature) | Very high (> 200 ºC) |

**Table S1.** Comparison with other techniques. Compared with other methods used for homostructure fabrication, we evaluate several techniques with our new method on key parameters of the fabrication process.

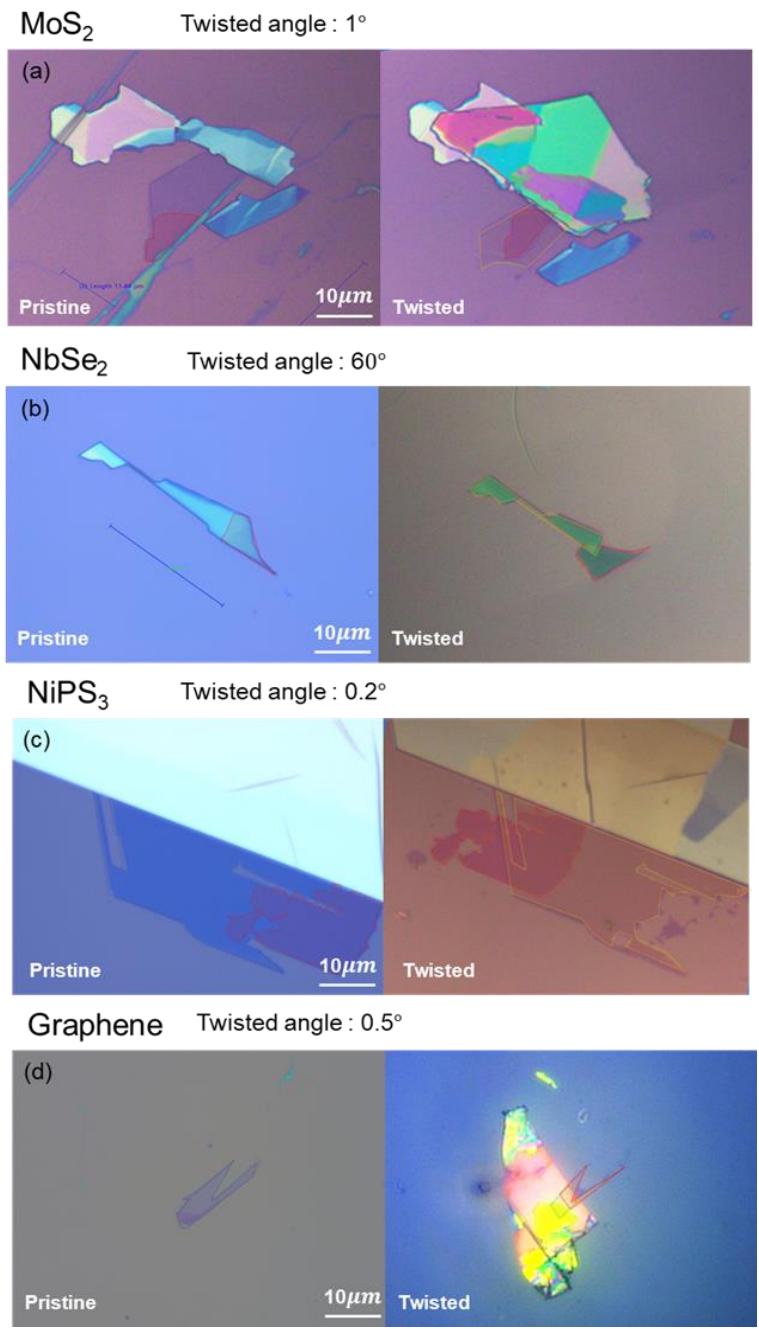

**Figure S2.** Other van der Waals twisted samples. (a-d) The images on the left were taken from pristine samples before the PCL tear and stack process, and the images on the right were obtained from the twisted devices. (a) Target sample: $MoS_2$ 1L protecting layer: $ZnPS_3$. (b) Target sample: few-layer $NbSe_2$, protecting layer: h-BN. (c) Target sample: $NiPS_3$, protecting layer: $CrPS_4$, (d) Target sample: Graphene, protecting layer: $ZnPS_3$. The scale bar is 10 μm.

**Supplementary Note 2: Additional STEM data**

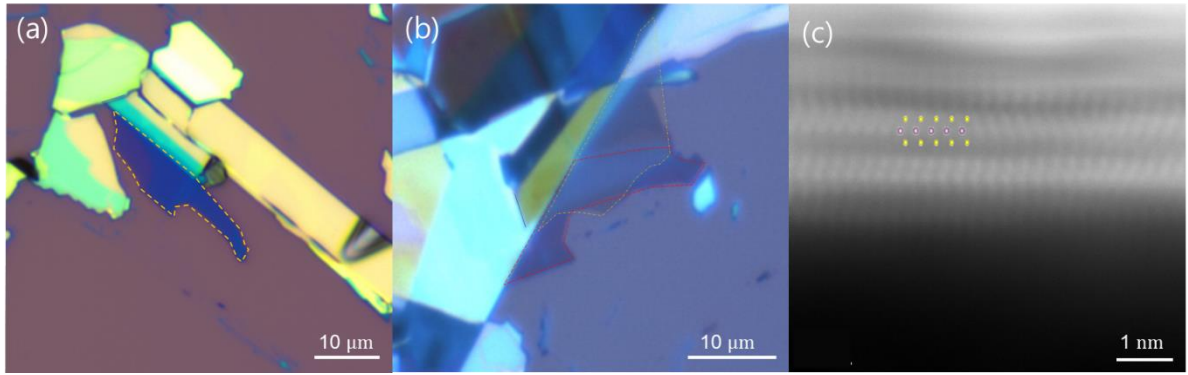

**Figure S3.** Few layers of MoS$_2$ STEM data. (a) Pristine 3L MoS$_2$ image. (b) Twisted MoS$_2$ with 30°. The scale bar is 10 μm. (c) STEM image of twisted MoS$_2$. The scale bar is 1 nm.

**Supplementary Note 3: Sample images used in measurements.**

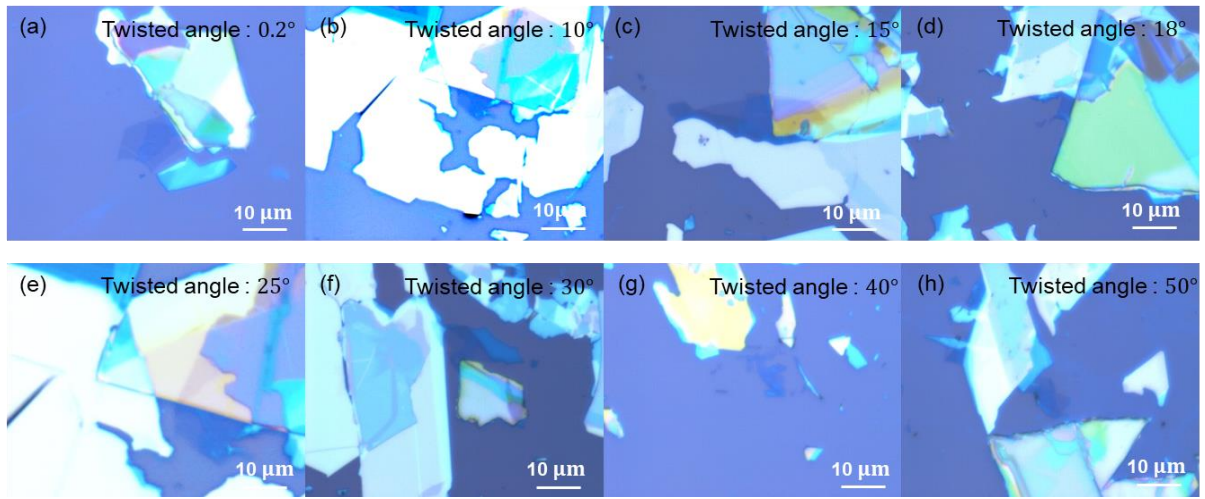

**Figure S4.** Twisted MoS$_2$ optical microscope images. (a-h) The samples are presented in the following order: 0.2°, 10°, 15°, 18°, 25°, 30°, 40°, and 50°. The scale bar is 10 μm.

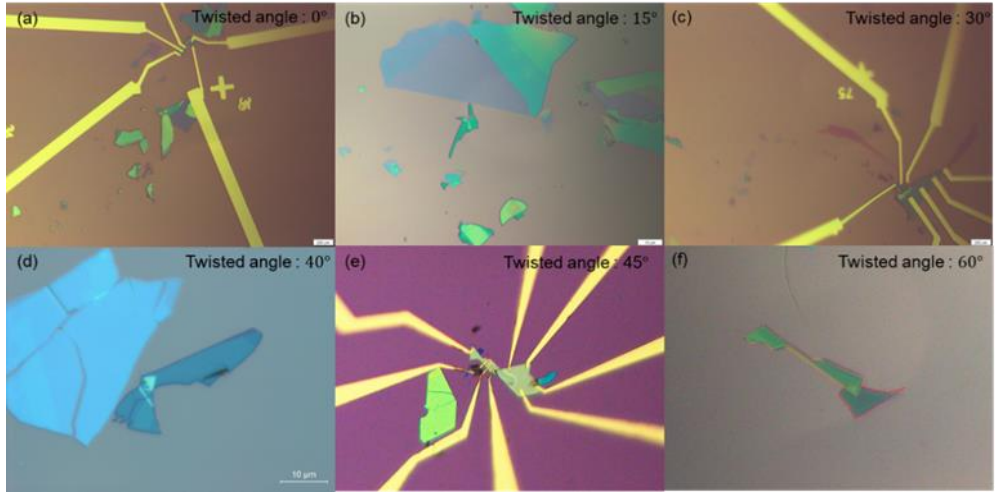

**Figure S5.** Twisted NbSe$_2$ optical microscope images. (a-f) The samples are presented in the following order: 0°, 15°, 30°, 40°, 45°, and 60°. The scale bar is 10 μm.

**Supplementary Note 4: Large-scale twisted homostructure**

For this test with large-scale twisted homostructure, we do not need a protective layer, unlike our smaller samples, as the cut line of the PCL stamp acts as a protective layer when handling large-sized samples, as shown below.

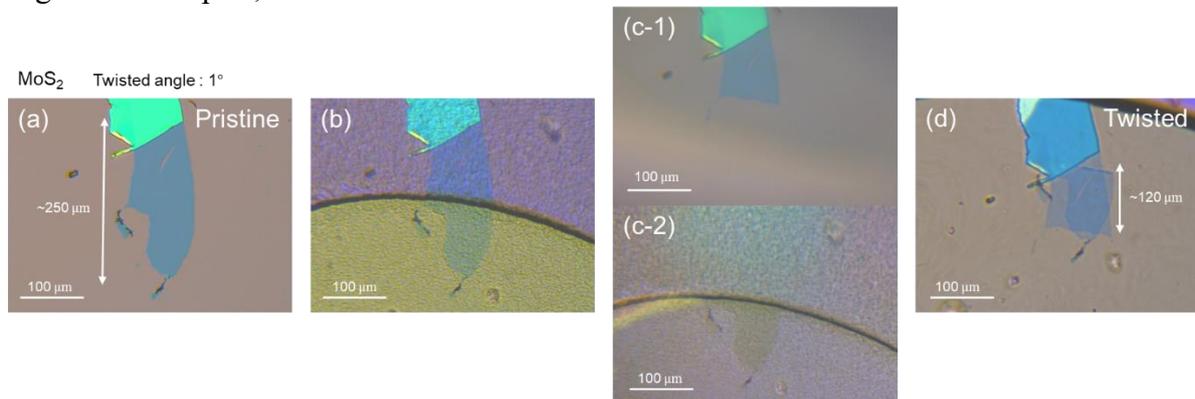

**Figure S6.** (a-d) Optical micrographs of successive steps. Target material: MoS$_2$, without a protective layer. (a) Finding target sample. (b) Align and attach the PCL stamp on the target layer. (c) Result of tear process. (c-1) Bottom layer on SiO$_2$ substrate, and (c-2) Top layer with PCL stamp. (d) Twist angle and stack process. The twisting angle is 1 °. The scale bar is 100 μm.

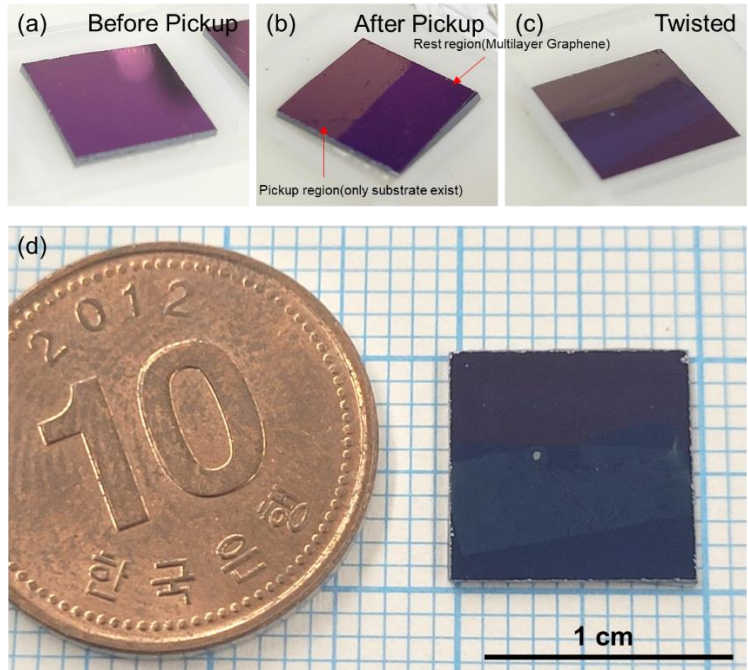

**Figure S7.** (a-d) Optical micrographs of successive steps. Target material: multilayer graphene on SiO$_2$(285 nm)/Si substrate, without a protective layer. (a) Preparing target sample. (b) Align and attach the PCL stamp on the target layer. (c) Result of tear and stack process. (d) Looking at this image, you can see that the size of the twisted part is similar to that of a Korean 10-won coin. This sample demonstrates that Lego structures can be built at an industrially viable scale while maintaining the most accurate angular control. The twisting angle is 13°. The scale bar is 1 cm.